# Quantum Secured Internet Transport

*Bernardo A. Huberman\*, Bob Lund\*, Jing Wang*
*\*CableLabs*

*Abstract*

Quantum computing represents an emerging threat to the public key infrastructure underlying transport layer security (TLS) widely used in the Internet. This paper describes how QKD symmetric keys can be used with TLS to provide quantum computing resistant security for existing internet applications. We also implement and test a general hybrid key delivery architecture with QKD over long distance fibers between secure sites, and wireless key distribution over short distance within each site Finally we show how this same capability can be extended to a TLS cipher scheme with "perfect security".

## INTRODUCTION

Quantum computing presents a unique challenge to current Internet security. The public key infrastructure (PKI) used to generate and distribute internet transport encryption keys is particularly vulnerable to quantum algorithms that provide an exponential speed-up in discovering private keys, thereby unlocking the symmetric encryption that protects data communications from eavesdroppers.

Internet encryption uses a key to encrypt data at the source and decrypt it at the destination. There are several ways for the source and destination to derive this symmetric key. One is to use a pre-shared key (PSK). Another way is to deploy a public key infrastructure (PKI), such as RSA, where the public and private key pairs at each site are used to derive a symmetric key. PKI is the dominant mechanism used today because PSK needs a secure way to share keys which is not available in the Internet. The security of public/private keys is fundamentally based on the premise that it is computationally infeasible to factor large numbers.

Unfortunately, factoring large numbers at an exponentially faster rate than on today's computers is one of the few algorithms that has been demonstrated on a quantum computer [1] [2] [3]. By snooping an encrypted internet session from the beginning of the PKI exchange through the end of data transmission, one could in principle, with the aid of a quantum computer, decipher the data – as it is sent or anytime in the future.

Fortunately, quantum key distribution (QKD) provides a means to address this challenge by offering a mechanism to generate *provably* secure symmetric keys over long distances. QKD is a new quantum technology that is starting to be tested by some academic and businesses institutions. It is part of a project to create a quantum network that will dramatically enhance Internet technology by enabling provably secure communication between any two points on Earth and beyond [4]. While several realizations, both in free space transmissions [5] and using fiber optics [6], have been reported in the literature, a number of issues need resolution in order to have a successful deployment. Current implementations, for example, require vertical integration of several complex technologies that impede its widespread adoption. Furthermore, the few cases where there has been an integration of QKD with existing applications are proprietary, further preventing its adoption.

We should mention that proposals are being made for the design of public key algorithms that will be resistant to quantum computers [7]. This so-called post-quantum security approaches will be in principle able to replace those vulnerable to quantum computing. While there are a number of proposed candidates, it will take at least a decade to decide on their robustness against attacks. In the meantime, data presently encrypted without QKD will be vulnerable in the future to decryption once quantum computers become available.

This paper describes a system design and implementation that integrates QKD with existing Internet services. This enables the utilization of standard protocols that can profit from the provably secure keys offered by QKD. Furthermore, it opens the door for internet service providers (ISP's) to offer quantum secured internet transport services, which in turn will create demand for the underlying optical transport services that will be needed in the future.

In the following we present a brief description of quantum key distribution, and discuss solutions to the issues that need to be resolved for QKD to be universally available for Internet services. Next we show how QKD keys can be used for Internet transport security and discuss a number of applications that create and use this secured transport. We then describe the nature of the interface between QKD secured transport and the underlying QKD networks, and provide examples of network services that need to be provided to facilitate the further growth of a usable quantum network. Following that we describe our implementation of a quantum secured transport layer and several experiments that we undertook to test and demonstrate its capabilities. We conclude by describing the use of QKD as a service in the context of metro and access networks.

## Quantum Key Distribution

Quantum key distribution (QKD) uses the exchange of quantum bits (qubits) between two parties to generate a symmetric key. The following table illustrates how QKD generates a shared secret key.

| Alice's random bit | 0 | 1 | 1 | 0 | 1 | 0 | 0 | 1 |
|---|---|---|---|---|---|---|---|---|
| Alice's random sending basis | + | + | × | + | × | × | × | + |
| Alice sending polarization | ↑ | → | ↘ | ↑ | ↘ | ↗ | ↗ | → |
| Bob's random measuring basis | + | × | × | × | + | × | + | + |
| Polarization Bob measures | ↑ | ↗ | ↘ | ↗ | → | ↗ | → | → |
| Shared secret key | 0 |  | 1 |  |  | 0 |  | 1 |

Source: https://en.wikipedia.org/wiki/Quantum_key_distribution

Alice generates a random classical bit string and randomly chooses one of two quantum basis (rectilinear and diagonal in this example) to transmit each bit. She transmits each qubit by polarizing a photon according to the classical bit value and the chosen quantum basis. Bob then measures each received photon by randomly using one of two agreed upon basis. Notice that Bob will measure a random value when he uses a different basis than Alice used to transmit. Alice publicly discloses the transmitting basis she used for each bit. Bob and Alice now share a subset of bits in the case where the transmitting and measurement bases are the same.

Bob and Alice can then detect errors or eavesdropping by comparing over a classical channel the measured values of a subset of the bits where the transmission and measurement bases are the same. Depending on the QKD protocol used, it has been proven that the remaining subset of shared bits cannot be observed by an eavesdropper; thus, making it *provably* secure [8]. Reference [9] provides a detailed description of the error detection and privacy amplification methods that yield a provably secure key.

This scheme can be instantiated in a network as shown in Fig. 1. As can be seen, QKD nodes *a* and *c* exchange keys via an optical fiber (or free-space optical link) and obtain a shared *key2*. *Key2* is provided to host computers at $IP_1$ and $IP_3$, respectively, where they are used to encrypt and decrypt an internet data exchange. The key exchange and data transmission can then take place over existing, classical fiber-optic networks.

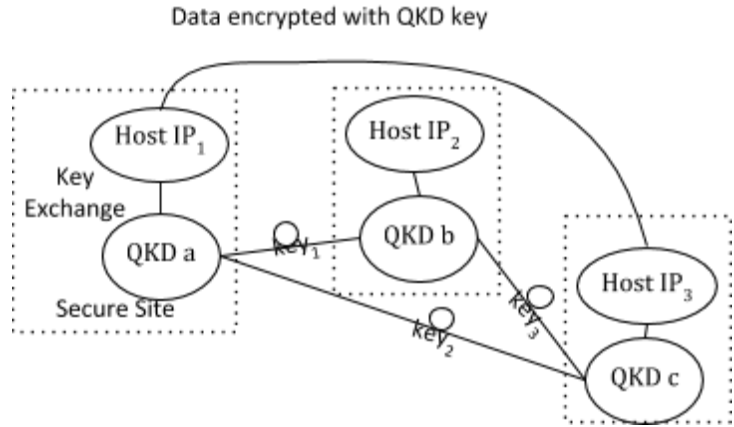

**Fig. 1 - Quantum Key Distribution**

## QKD FOR THE INTERNET

### QKD and Internet Transport Security

Transport Layer Security (TLS) [10] and the deprecated Secure Sockets Layer (SSL) protocols [11] are the basis for web services transport security used by the vast majority of web services. Pre-shared keys can be used by TLS, as shown in Fig. 2.

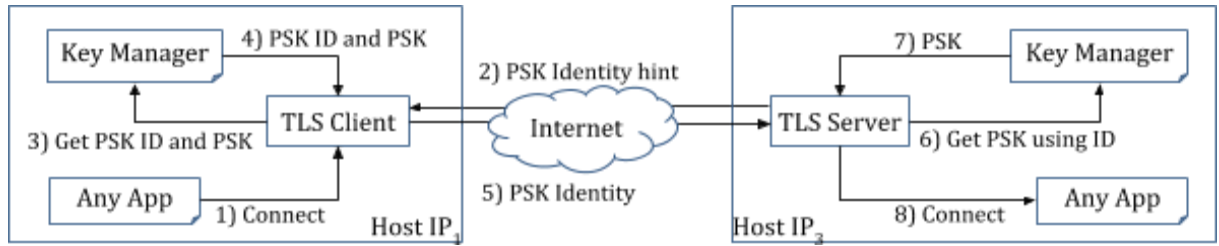

**Fig. 2 - TLS Pre-shared Keys**

While TLS is widely used by internet and web services, several issues prevent the immediate adoption of QKD for PSKs. Since there is no secure way to pre-share keys, most TLS implementations don't support PSK. Even in cases where it is supported, there is no standard method for QKD keys to be used for PSK TLS. Furthermore, there is no simple mechanism to distribute QKD PSKs to applications. It is because of these reasons that applications aren't written to make use of PSKs.

The next section describes a method for using QKD with TLS, to be followed by a description of how QKD TLS can be made available to applications that don't natively support PSK TLS.

### QKD PSK with TLS

Reference [12] defines how generic PSKs are used in TLS. The fundamental issue is how does a client and server agree on a common PSK when either or both have multiple PSKs that might be used. This is addressed by requiring the server to provide a "PSK identity hint" in the *ServerKeyExchange* message sent during the establishment of the TLS session. With this, the client can then select a PSK corresponding to the identity hint. The client then includes the "PSK identity" back to the server in the *ClientKeyExchange* message that the server uses to select the same PSK used by client.

Since no mechanism has been defined to associate QKD keys with "PSK identity hint" and "PSK identity" information we decided to use the following method.

The European Standard Organization ETSI defines a framework for QKD networks to make shared quantum keys available between a master secure application entity (SAE) and a slave SAE [13]. We define the operation of the ETSI method in the context of TLS as follows:

1. The master and slave SAEs represent the TLS client and server, respectively.
2. The TLS client and server network addresses (either IP or fully qualified domain name) are used as the respective SAE identifier. This implies that the "Host IPx" in Fig. 1 corresponds to master or slave SAEs.
3. The "QKD x" in Fig. 1 corresponds to the key management entity (KME) in [13].
4. The key exchange in Fig. 1 corresponds to the "Protocol Specifications" section in [13] (see Fig. 3).
5. The TLS server network address is used as the "PSK identity hint".
6. The following JSON object is used as the "PSK identity": { keyId: keyInfo.keyId, clientId: client.addr }, where keyId is the identity of the PSK selected by the client and client.addr is the TLS client network address.
7. The key manager represents the key exchange functionality in Fig. 1 "Host IPx".

With the above, the steps shown in Fig. 1 and Fig. 2 were extended as follows:

**Fig. 2 step 3):** The TLS client invokes the key manager with the "PSK Identity hint", i.e. the slave SAE network address.

**Fig. 1 key exchange):** The key manager requests PSKs with corresponding PSK Identities using the "Get key" section in [7] where the SAE field is the "PSK Identity hint".

**Fig. 2 steps 4 and 5):** The key manager selects PSK and PSK Identity and returns the JSON object described above as the "PSK Identity" to the TLS client.

**Fig. 2 step 6):** The key manager requests PSKs using the "Get keys with key ids" section in [13] where the SAE and the key_ID fields are set to clientId and keyId properties in the JSON described above.

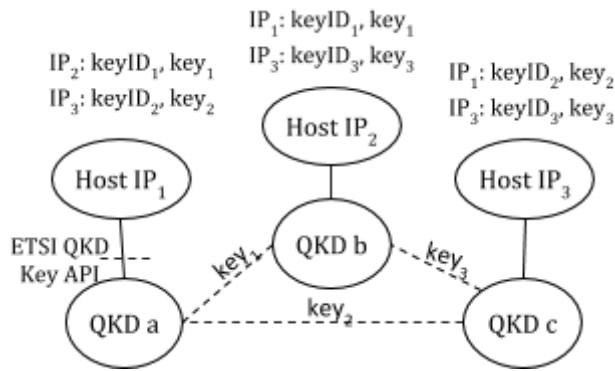

**Fig. 3 - QKD Layer Key Management**

An essential element in the design was a key manager implementation that prefetched keys from the QKD layer (ETSI QKD Key API Fig. 3) that are then available on-demand from TLS (steps 3 and 6 in Fig. 2). This ensures that any additional latency introduced by QKD is not seen in key use by TLS.

## Application Support for QKD-TLS

The method described in the previous section enables the utilization of shared keys from a QKD network in applications that use TLS. However, as noted earlier, the use of pre-shared keys is not widely supported because there is no established means for distributing them. Furthermore, there is no agreement on how keys from a QKD network would be made available to applications.

However, the concept of a proxy, widely used in the Internet, can be exploited to expose TLS QKD PSK to existing applications in a manner transparent to those applications (Fig. 4).

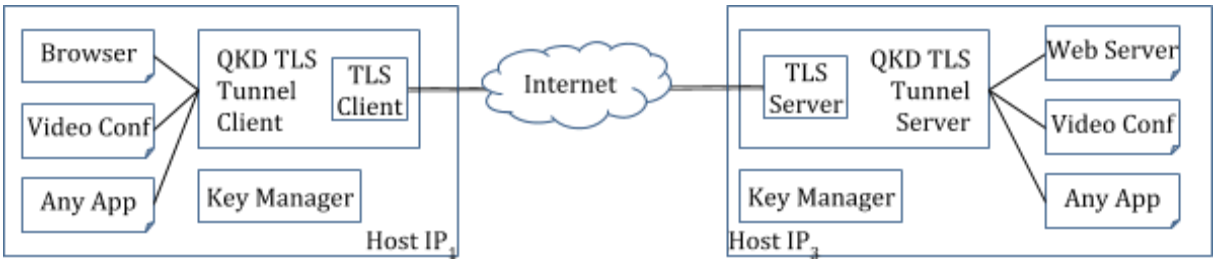

**Fig. 4 - QKD TLS Tunnel**

Most computing platforms provide a system-wide means to redirect classes of network traffic to specified TCP/UDP ports. Specialized applications, such as proxies, then route that traffic as appropriate.

The QKD TLS tunnel client and server are new proxy applications that implement Fig. 2, along with the extensions defined in this paper. Network traffic from existing applications then connect to the tunnel client and server, which communicate with each other over a QKD-TLS tunnel setup over the Internet. In this way, the existing, unmodified applications take advantage of QKD TLS across the Internet.

# A QKD TLS Proof-of-Concept System

A proof of concept (POC) of the QKD TLS system described in the previous sections was developed at CableLabs. Fig. 5 shows a vision of future communication networks protected by QKD. Since current technology only allows QKD using photons, two communication parties need to be connected by optical fiber or free space optics links. Since it is impossible however to connect all devices with fibers, the ubiquity and flexibility of wireless access makes it the best choice for the key distribution within the last segment of networks.

We describe a hybrid key delivery architecture including QKD over long distance fibers between secure sites, and wireless key distribution over short distance within each site (Fig. 5). These secure sites, which could be bank buildings, business campuses, or government offices, are connected by QKD links via optical fibers. Within each site, the keys are delivered to mobile devices wirelessly. Although the wireless links still use conventional cryptography, there are other methods to make them safe. For examples, free-space-optics, visible light communication, and directional millimeter wave beams, cannot penetrate walls and can be confined within the secure site. While in legacy networks the whole communication link must be secured from end to end, in this architecture, only the secure sites need to be protected.

This hybrid architecture is a promising candidate for a first phase deployment of security-as-a-service. Different granularities of securities are realized, e.g., absolute security over long distance between secure sites and computational security over short distance within each site, thus trading security for mobility and flexibility.

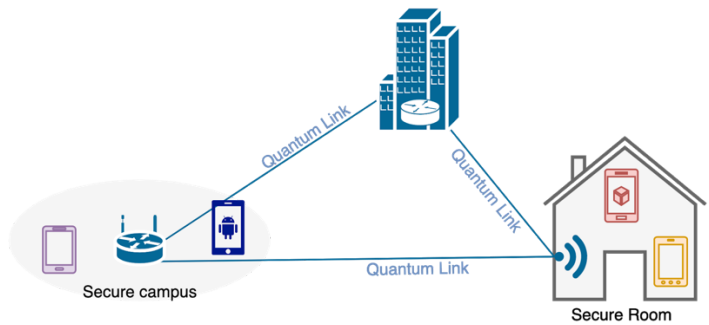

Fig. 5 - QKD-TLS Proof of Concept

The experimental setup is shown in Fig. 6. We deployed a QKD system and used its keys in the TLS tunnel to protect the wireless communication between two mobile users.

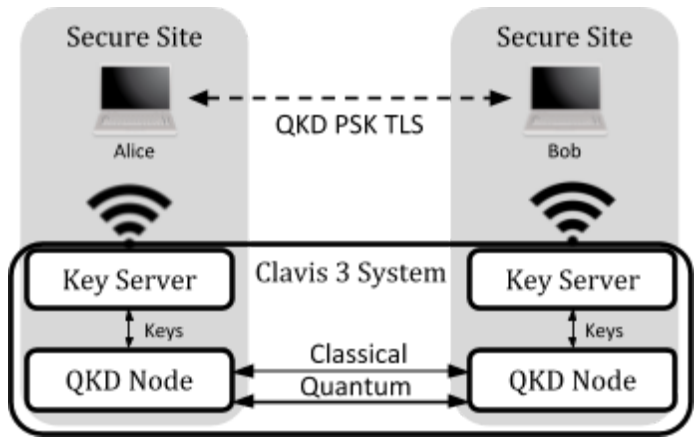

Fig. 6 - Experimental Setup

Two secure sites were separated by 25 km and connected by a QKD quantum channel. Besides the quantum channel, there was also a classical channel carrying traditional traffic between the two sites and serving as an auxiliary channel for QKD. The QKD-TLS tunnel was implemented using NodeJS v13.6 [14]. NodeJS is a widely deployed web services platform with v13.6 being the first version that implements PSK TLS. A key management layer implementing the ETSI QKD key management API was also developed.

The quantum layer is based on the ID Quantique Clavis 3 QKD system [15], which includes truly random bit generation and quantum key distribution. After the key exchange, both sites stored the keys in the key servers. Mobile users can then fetch the keys from the servers and use them to protect their communications. Thanks to the versatility of TLS, we were able to demonstrate how internet traffic, such as simple web browsing, video streaming and WebRTC video and text chat, could be transported over the QKD-TLS tunnel. Furthermore, after obtaining the keys, the users could roam away from the secure sites and continue their secure communication protected by their stored keys. Once the keys are consumed, mobile users could to return to the secure sites to fetch new keys.

## QKD Networks As A Service

Deploying a QKD network, such as the one illustrated in Fig. 3, requires a dedicated fiber link among nodes. Furthermore, since quantum mechanics precludes any amplifier or repeater within a quantum link, QKD deployments will be limited to ≈ 100 miles[1] in the near term if one does not want to resort to teleportation techniques or quantum routers [16]. But this is not an obstacle for the emergence of ISP metro QKD services that can provide on-demand point-to-point optical connections with the required fidelity via wavelength switches or reconfigurable optical add/drop multiplexers (ROADM) [17].

This metro service could be further expanded by using novel ways of delivering quantum keys over the optical access network. Specifically, we are now working on integrating QKD into passive optical networks (PON), which will enable the delivery of keys to end users exploiting the fiber deep architecture of cable networks. This will also allow for the reuse of classical infrastructure for QKD, with the ensuing reduction in system costs (Fig. 7).

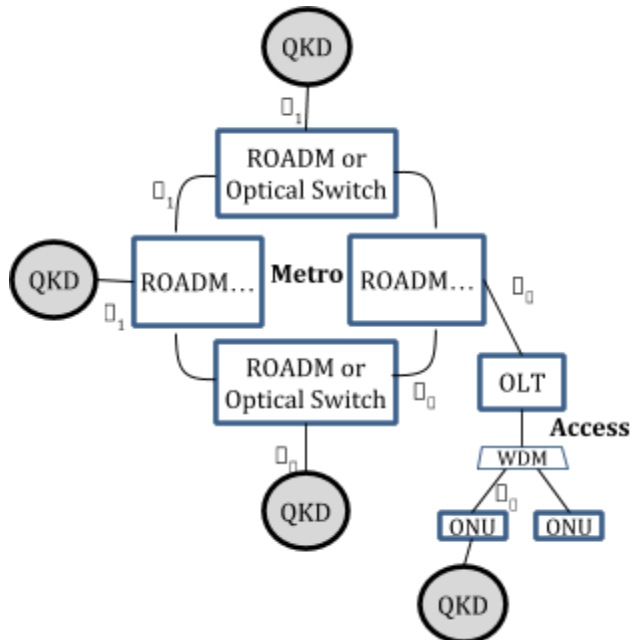

Fig. 7 - Metro QKD Optical Transport

Longer range qubit transmission is possible but requires using quantum entanglement and repeaters. Adding a control plane results in a quantum router which enables a true quantum Internet. We have already shown how such a router would operate. Fig. 8 illustrates the components of a quantum router. More details can be found in [16].

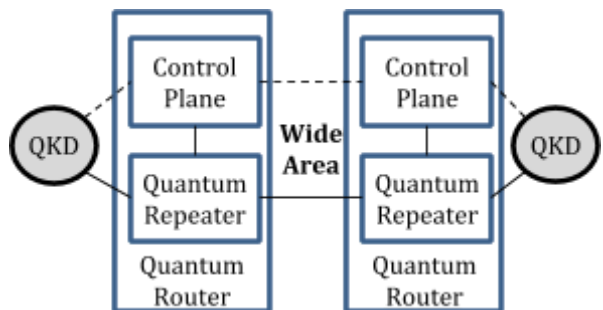

Fig. 8 - Quantum Internet Router

---

[1] QKD has the notion of a QKD trusted repeater that chains together QKD point-to-point links, thereby overcoming the optical limit. A 3rd party trusted repeater seems at odds with security requirements necessitating QKD in the first place.

# PERFECT SECURITY WITH QKD

Using QKD pre-shared keys instead of symmetric keys derived from PKI makes TLS much more resistant to quantum computers exploiting Shor's algorithm. However, the underlying symmetric encryption algorithm used, AES for example, is still only computationally secure. Perfect, i.e. provable, encryption security can be achieved by using the one-time-pad (OTP) encryption with QKD pre-shared keys.

One-time-pad is an encryption technique that uses a one-time pre-shared key the of the same length as the message to be sent [18]. Each character of the message to be sent is encrypted via some operation with a character in the key; decryption is done with a reverse of the operation using the same key character on the encrypted message character. OTP has a long history of use with pre-shared keys consisting of written text. Gilbert Vernam was issued a patent for the method of encrypting and decrypting a digital message with a digital pre-shared key using the exclusive or (XOR) operation [19]. Claude Shannon later introduced the concept of "perfect security" and proved that the OTP with a single-use, completely random key provided perfect security [20]. Generating a perfectly random key as long as a message which can only be used once, and sharing it securely has been a significant obstacle to using OTP in practice, however.

As noted earlier in this paper, QKD directly solves the problem of securely sharing a key in a provably secure manner. A perfectly random shared key can be generated in one of two ways. A number of commercially available quantum mechanisms exist for Alice to create a random key [21]. Alice can then share this key via QKD. A second method is for Alice to create two strings of qubits in the |0> state, entangle each pair of qubits (one from each string) and then share one string of entangled qubits with Bob via QKD. If the proper quantum entanglement circuit is used, Alice and Bob can each measure their entangled string which will result in a correlated random string [22].

QKD key distribution shown in Fig. 3 can be used to create random PSK for OTP. A perfectly random OTP of any length can be formed by using the shared sequence of (keyID1, key1), … (keyIDn, keyn). Alice uses her copy of the OTP to encrypt a message, then sends the sequence of keyIDs to Bob, who uses the associated sequence of keys, to decrypt the message.

Using a OTP in the QKD TLS system design in Fig. 4 requires that TLS be extended to implement the OTP XOR algorithm, which is straightforward. It also requires that the TLS client be able to use the sequence of *keys* to encrypt the message and to send the associated sequence of *keyIDs* to the TLS server so it can decrypt the message; this is not straightforward to implement. Since the OTP XOR operation is simple, an interim approach would be to replace the TLS client and server with a specialized OTP client and server that share a signaling channel to exchange the sequence of *keyIDs.* The application interface to the QKD tunnel client and server is the same in either case so one could start with the interim solution and migrate to the OTP integrated in TLS solution.

# CONCLUSIONS

In this paper we described how QKD symmetric keys can be used with TLS to provide quantum computing resistant security for existing internet applications. We also implemented and tested a general approach for wirelessly distributing QKD keys within secure sites. Finally, we showed how the QKD-TLS tunnels can evolve to use quantum key based one-time-pad to provide perfectly secure internet transport.